\documentclass[twocolumn,pra]{revtex4-1}
\usepackage{amsmath,bbm,graphicx,hyperref}

\newcommand{\GM}{\Gamma_{\text{max}}}

\newcommand{\Gup}{\Gamma_{\uparrow}}
\newcommand{\Gdn}{\Gamma_{\downarrow}}
\newcommand{\moldense}{\rho_0}
\newcommand{\Gupint}{\Gamma_{\uparrow}^{\text{int}}}
\newcommand{\lho}{\ell_{\text{HO}}}
\newcommand{\rhohat}{\hat{\rho}}
\newcommand{\wzpl}{\omega_{\text{ZPL}}}
\renewcommand{\vec}[1]{\mathbf{#1}}

\begin{document}

\title{Spatial dynamics, thermalization, and gain clamping in a photon
  condensate}

\author{Jonathan Keeling}
\affiliation{SUPA, School of Physics and Astronomy, University of St Andrews, St Andrews, KY16 9SS, United Kingdom}
\author{Peter Kirton}
\affiliation{SUPA, School of Physics and Astronomy, University of St Andrews, St Andrews, KY16 9SS, United Kingdom}

\date{\today}

\begin{abstract}
  We study theoretically the effects of pump-spot size and location on
  photon condensates.  By exploring the
  inhomogeneous molecular excitation fraction, we make clear the
  relation between spatial equilibration, gain clamping and
  thermalization in a photon condensate.  This provides a simple
  understanding of several recent experimental results. We  find
  that as thermalization breaks down, gain clamping is imperfect,
  leading to ``transverse spatial hole burning'' and multimode condensation.  This opens
  the possibility of engineering the gain profile to control the condensate
  structure.
\end{abstract}

\maketitle

\section{Introduction}
\label{sec:introduction}

The laser has long served as a prototype for phase transitions in
driven dissipative systems~\cite{Graham1970a,haken75}.  While for a
single-mode cavity the transition is mean-field-like, in a multimode
cavity~\cite{staliunas} spatial fluctuations are possible, enabling
non-trivial critical behavior.  In the last decade, there has been
much interest in other examples of phase transitions in
driven-dissipative systems.  In part, this has been motivated by
experiments on polariton
condensation~\cite{Kasprzak2006,Balili2007,Carusotto2012}.  However,
there are also experiments on cold atoms in optical 
cavities~\cite{Baumann2010,Baden2014}, and proposals for experiments in
``coupled cavity arrays''~\cite{Angelakis2007a,Hartmann2006,Greentree2006} 
of superconducting
circuits~\cite{Underwood2012} or hybrid quantum
systems~\cite{Zou2014a}.  There are also intriguing connections
between the dynamics of these quantum systems, and the study of
similar questions on the dynamics of classical ``active
matter''~\cite{Marchetti2013}, as studied in photo-excited colloidal
systems~\cite{Palacci2013}.  These systems all address common
questions of how a flow of energy through the system affects the
collective dynamics of the system, and the
emergence of spatial structures.

Closely related to both polariton condensates and photon lasers are
experiments on Bose-Einstein condensation (BEC) of
photons~\cite{Klaers2010c} in organic-dye-filled microcavities. Unlike
polaritons, these systems have no strong matter-light coupling and so
the normal modes are non-interacting photons.  However thermalization
is possible~\cite{Klaers2010b} via the dye molecules. If a photon can
be absorbed and emitted many times before it leaves the cavity, the
photon gas achieves thermal equilibrium with the dye.  Thus, by
adjusting the rates of absorption and emission, or cavity decay, one
may interpolate between an equilibrium BEC, and a strongly dissipative
dye laser~\cite{Schafer1990}.  We will refer to condensation
throughout this paper, but we present phenomena that can be
interpreted either as lasing or BEC\@.  Following these experiments
many theoretical
works~\cite{Klaers2012a,Sob'yanin2012,Snoke2012,Kirton2013b,Kruchkov2014,VanderWurff2014,DeLeeuw2014,Chiocchetta2014,Nyman2014,Sela2014,Kirton2015,DeLeeuw2014a,Chiocchetta2015}
explored topics including equilibration, phase coherence, and photon
statistics of the photon BEC, and later experiments studied photon
statistics~\cite{Schmitt2014}.  

Recently, two experiments~\cite{Marelic2015,Schmitt2014b} studied the
spatial profile and dynamics of the photon BEC and their dependence on
pump-spot size and location, observing behavior beyond the scope of
existing models.  Spatial profiles below threshold were also
previously studied in~\cite{Klaers2010b}.  These works motivate this
paper.  Studying spatially varying systems moves away from the domain
of simple ``mean-field'' models of lasing or phase transitions:
spatial modes allow for non-mean field critical behavior at phase
transitions, and for spatial decay of coherence.  This has been
explored experimentally for polaritons in one-~\cite{wertz10} and
two-dimensions~\cite{Roumpos2012}.  Considering such critical
behavior in extended systems, theoretical work has shown that
features beyond the equilibrium classification~\cite{Hohenberg1977}
can arise, such as new critical exponents in three
dimensions~\cite{Sieberer2013}, the destruction of algebraic order in
two-dimensional systems~\cite{Altman2015}, and potential novel
universality classes in one dimension~\cite{Marino2015}.  Multimode
cavity systems --- i.e.\ spatially extended systems --- also allow
for transverse pattern formation, as has been studied in
lasers~\cite{staliunas}, and for
polaritons~\cite{Manni2011,Tosi2012,Nelsen2013,Gao2015}. Very
recently, there has been an experimental realization of a system of
cold atoms in a multimode optical cavity~\cite{Kollar2014}.  An
important distinction exists between such atomic experiments, where
photons couple to density or spin waves of the atoms, with the atom
number being conserved, vs exciton-polaritons where photons couple to
the exciton itself, creating or destroying excitons.  Nonetheless,
these systems provide an additional complimentary perspective on the
physics of driven-dissipative matter-light systems.

The aim of this paper is to introduce a model capable of describing
how the size and shape of the pump profile affects the spatial
profile of a photon BEC\@. 
In order to describe the spatial profile of a condensate, a widely
used approach is to derive order parameter equations, i.e.\ a
partial differential equation determining the time evolution of a
field $\Psi(\vec r)$, representing the condensate order parameter.
The equations determining the spatial profile of a condensate are
distinct for closed (conservative) and open (dissipative) systems.  In
a closed BEC, this equation is the Gross-Pitaevskii
equation~\cite{pitaevskii03} (GPE), which can be written in the form
$i \hbar \partial_t \psi = \delta E[\Psi]/\delta \Psi^\ast$.  Such an
equation conserves an energy functional $E[\Psi]$, with corresponding
phase evolution of the order parameter.  In contrast, for a purely
dissipative system, the time dependent Ginzburg Landau
equation~\cite{Ginzburg} (GLE) describes irreversible relaxation,
$\partial_t \psi = - \Gamma \delta E[\Psi]/\delta \Psi^\ast$, so that
the final state is a state of minimum energy.  
A classification of such order parameter equations has been given
by \citet{Hohenberg1977}.  
Combining both conservative and dissipative terms leads
to the complex GPE or GLE~\cite{Aranson2002a}, widely used for
polariton condensates~\cite{wouters07,Keeling2008a,wouters08}.
In some cases, such equations can show critical behavior
outside the Hohenberg--Halperin classification~\cite{Sieberer2013}.
Similar equations also arise in nonlinear optics~\cite{staliunas},
where dispersive shifts (i.e.\ nonlinear dielectric functions,
depending on the field amplitude $|\Psi|^2$) compete with dissipative
terms describing loss and gain;
for example, in a class-A laser, the dynamics of the gain medium
can be adiabatically eliminated, leading to a complex Ginzburg--Landau
equation of motion for the field amplitude~\cite{Graham1970a}.

In this paper we make use of a different approach, considering
density matrix equations of motion, rather than order
parameter equations.  This is because
 order parameter equations generally only crudely model relaxation
to a thermal state --- in particular, the order parameter normally
only describes the macroscopically occupied mode(s), and neglects
thermal fluctuations.  For many examples of pattern formation in
nonlinear optics this is entirely appropriate: no thermalization
occurs, and this is accurately reproduced by the order parameter
equation.  
However, for photon BEC, thermalization is a key feature of
the observed behavior, and so a complete model should be able to
explain how this interacts with spatial pattern formation.  Extensions
of order parameter equations to include energy dependent gain rates
have been developed to address this for
polaritons~\cite{Wouters2010b,Wouters2012a}.  By including also noise terms
phenomenologically, these can yield thermal distributions.  In this
paper we instead follow an approach that proceeds directly
from our microscopic model~\cite{Kirton2013b,Kirton2015}.  We show that
for weak coupling one can derive a tractable model combining spatial
dynamics with energy relaxation.  This model describes how the spatial
profile is determined by the competition between energy relaxation and
loss, and can explain the recent
experiments~\cite{Marelic2015,Schmitt2014b}.

The remainder of this paper is organized as follows.
Section~\ref{sec:model} describes our model of the experiments, and
derives a master equation for the photon modes and
electronic state of the molecules, eliminating the fast dynamics of
molecular vibrations.  From this model, we derive coupled equations
for the population of excited molecules, and the populations and
coherence of photon modes.  Using these equations,
section~\ref{sec:steady-state} discusses the steady state properties
of the photon cloud.  In Sec.~\ref{sec:far-below-threshold} we first
show how, far below threshold, the occupation of photon modes depends
both on their energy (controlling the rate of emission and absorption
for that mode), and also the overlap between the photon mode and the
profile of the pump.  This behavior occurs at weak pumping because
there the excitation profile of molecules follows that of the pump.
The same approach allows us to understand how the pump profile
affects the threshold power required for condensation, discussed in
Sec.~\ref{sec:behav-at-thresh}.  Once above threshold, the profile of
excited molecules is significantly modified by the condensed photons,
via a kind of transverse spatial hole burning.  We discuss the
consequences of this in Sec.~\ref{sec:above-thresh-transv}.  In
Section~\ref{sec:dynamics} we then turn to study the early-time
transient dynamics of a condensate following an off-center pump.  We
show how the spatial oscillations evolve due to reabsorption of cavity
light, leading ultimately to thermalization.  Finally,
section~\ref{sec:conclusion} provides some conclusions and outlook
from our work.

\section{Model}
\label{sec:model}

\begin{figure}[thpb]
  \centering
  \includegraphics[width=3.2in]{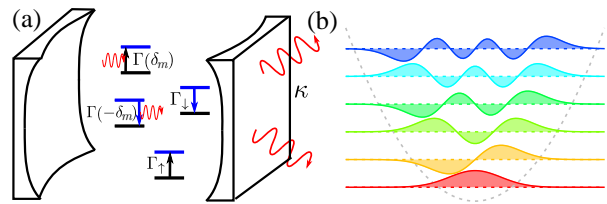}
  \caption{(Color online) (a) Cartoon of model system: molecules are
    represented by two electronic states (HOMO and LUMO levels),
    dressed by ro-vibrational excitations.  (b) Gauss-Hermite
    eigenfunctions of an harmonic oscillator.  }
  \label{fig:cartoon}
\end{figure}

The photon BEC system consists of dye molecules coupled to photon
modes in an optical cavity (see Fig.~\ref{fig:cartoon}(a).  Each molecule
has a complex optical spectra, due to the ro-vibrational dressing of
the electronic spectrum.  Despite this, one can nonetheless consider
only two electronic states, the highest occupied molecular orbital
(HOMO) and lowest unoccupied molecular orbital (LUMO).  Each of these
levels is however dressed by ladder(s) of rotational and vibrational
excitations of the molecules.  As discussed in our previous
work~\cite{Kirton2013b,Kirton2015}, one may adiabatically eliminate the
vibrational states, leading to absorption and emission rates
$\Gamma(\pm \delta)$ for photon modes detuned by
$\delta=\omega-\wzpl$ from the Zero Phonon Line (ZPL) of the
molecule.  This results in a model in which the electronic state (HOMO
or LUMO) of each molecule is explicitly represented, while the effects
of the ro-vibrational excitations appear implicitly in the
structure of the rates $\Gamma(\pm \delta)$ discussed further below.

To incorporate inhomogeneous pumping we must consider the overlap
$\psi_m(\vec{r}_i)$ between the transverse mode function of photon
mode $m$ and a molecule at $\vec{r}_i$. 
We do not include here effects of the longitudinal mode profile, 
as we consider  cases where only a single longitudinal
mode is relevant (i.e.\ close enough to resonance with the gain
medium).  In these cases, except for very high order modes, the longitudinal
mode profile does not vary significantly between the modes, and so any
effects of overlap between the longitudinal mode profile and the
excited molecules can be absorbed into a constant factor in the
definition of emission and absorption rates.  
For the transverse modes, curvature of the cavity
mirrors leads to an in-plane harmonic trap, so that $\psi_m(\vec{r})$
are Gauss-Hermite functions (see Fig.~\ref{fig:cartoon}(b)) in two
dimensions and the corresponding frequencies are harmonically spaced,
$\omega_m=\omega_c+(m_x+m_y)\epsilon$, where $m$ combines both $m_x$
and $m_y$ indices. The ``cavity cutoff'' $\omega_c$ is set by the
cavity length.  We write the master equation describing the system as
two terms, $\partial_t \rhohat = \mathcal{M}_0[\rhohat] +
\mathcal{M}_\text{int}[\rhohat]$. The bare part is:
\begin{multline}
  \label{eq:1}
  \mathcal{M}_0[\rhohat]
  =  -i \sum_m \left[ \omega_m \hat{a}^\dagger_m \hat{a}^{}_m, \rhohat\right]
  + \sum_m \frac{\kappa}{2}  \mathcal{L}[\hat{a}_m,\rhohat] 
  \\+
  \sum_i  
  \frac{\Gup(\vec{r}_i)}{2}\mathcal{L}[\hat{\sigma}^+_i,\rhohat] 
  +
  \sum_i  
  \frac{\Gdn}{2} \mathcal{L}[\hat{\sigma}^-_i,\rhohat],
\end{multline}
where $\mathcal{L}[\hat X,\rhohat]=2 \hat X^\dagger \rhohat \hat X -[\hat
X^\dagger \hat X,\rhohat]_{+}$.  The operator $\hat{a}^\dagger_m$ creates
a photon in mode $m$, and we assume all modes have  decay rate
$\kappa$. The electronic state of  molecule $i$ is represented
by Pauli operators $\hat\sigma_i^{x,y,z}$.  In addition to 
coupling to the cavity (see below), each molecule  has a pumping 
rate $\Gup(\vec{r}_i)$, and a non-cavity decay rate $\Gdn$ incorporating
fluorescence into all modes other than the confined cavity modes.
Other than the inhomogeneous pump, $\mathcal{M}_0[\rhohat]$ matches
Refs.~\cite{Kirton2013b,Kirton2015}.

The term $\mathcal{M}_{\text{int}}[\rhohat]$, describing molecule-photon interaction, can be treated at
various levels of approximation, according to whether we include
coherence between different photons modes.  Including such inter-mode
coherence is numerically expensive, and is only necessary when
significant coherence exists. The numerical cost arises because, if we
truncate the equations to consider $N_m$ photon modes, the full
coherence matrix scales as $N_m^2$.  As discussed later, in order to
keep all significantly populated modes (when $k_B T \gg \hbar
\epsilon$), we need relatively large values of $N_m$.  Thus, with the
full equations, it is only feasible to simulate a few hundred
picoseconds of time evolution, far shorter than the timescale required
to reach the steady state. In what follows, we therefore first present
the full equations of motion, used to study transient dynamics, and
then introduce the ``diagonal approximation'', providing a more efficient
approach when inter-mode coherence can be neglected.

\subsection{Fully coherent model}
\label{sec:fully-coher-equat}

\begin{figure}[thpb]
  \centering
  \includegraphics[width=3.2in]{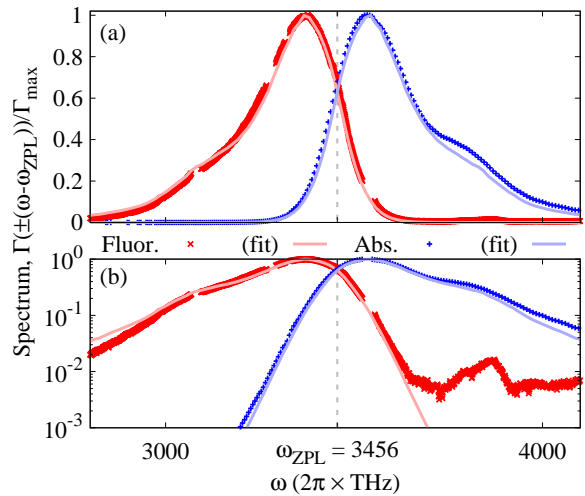}
  \caption{(Color online) (a) Absorption and fluorescence spectrum of
    Rhodamine 6G on a (a) linear or (b) logarithmic scale. Points are
    experimental data~\cite{nyman_spectrum} (for dye in Ethylene
    Glycol), lines show the fits $\Gamma(\pm\delta)$ satisfying
    $\Gamma(\delta)=\Gamma(-\delta)e^{\beta \delta}$ at room
    temperature.  Here and throughout, we plot angular frequency,
    $\omega=2\pi f$.  To indicate this we write the units as
    $2\pi$THz. }
  \label{fig:spec}
\end{figure}

We denote the most complete form of the molecule-photon interaction 
 $
\mathcal{M}_{\text{int}}^{}[\rhohat] =
\mathcal{M}_{\text{int}}^{\text{full}}[\rhohat]$,
which takes the form:
\begin{multline}
  \label{eq:2}
  \mathcal{M}^{\text{full}}_{\text{int}}[\rhohat]
  \equiv \!\!  \sum_{m,m^\prime,i} \!\!
  \psi^\ast_m(\vec{r}_i) \psi^{}_{m^\prime}(\vec{r}_i)
  \bigl\{
    K(\delta_{m^\prime})
    [ \hat{a}_{m^\prime} \hat{\sigma}_i^+ \rhohat, 
    \hat{a}_m^\dagger \hat{\sigma}^-_i]
    \\
    +
    K(-\delta_m) 
    [ \hat{a}^\dagger_m \hat{\sigma}_i^- \rhohat, 
    \hat{a}_{m^\prime} \hat{\sigma}_i^+] 
  \bigr\} + \text{H.c.}.
\end{multline}
The complex function $K(\pm\delta_m)$, discussed next, encodes the
molecular absorption (emission) rate vs.\ the detuning
$\delta_m = \omega_m - \wzpl$ between mode $m$ and the
molecular Zero Phonon Line (see dashed line in
Fig.~\ref{fig:spec}).

For simple molecules, $K(\delta)$ can be calculated
explicitly~\cite{Kirton2015}.  Alternatively, one may use
experimentally measured spectra $\Gamma(\delta)$, and find $K(\delta)$
by analytic continuation --- causality requires that $K(\delta)$ is
analytic in the lower half plane.  As noted
previously~\cite{Klaers2010b,Klaers2010c,Kirton2013b}, thermalization
of photons requires that $\Gamma(\delta)$ obeys the Kennard-Stepanov
(KS) relation~\cite{Kennard1918,Kennard1926,Stepanov1957},
$\Gamma(\delta)=\Gamma(-\delta) e^{\beta \delta}$.  We therefore use a
function $\Gamma(\delta)$, shown in Fig.~\ref{fig:spec}, that fits the
experimental spectra, while satisfying the KS relation.  The procedure
used is described in appendix~\ref{sec:extr-gamm-from}. This determines
$\Gamma(\delta)$ up to a prefactor.  We denote $\GM =
\text{max}[\Gamma(\delta)]$, and discuss below how to estimate $\GM$
from experimental results.  The function $K(\delta)$ is then found
by standard analytic continuation of $\Gamma(\delta)$ to the lower
half plane.

Note that while equation~(\ref{eq:2}) includes coherence between
photon modes, it neglects coherence between molecules.  This is
because dye and solvent molecules collide frequently causing rapid
dephasing.  Inter-mode coherence will be required to understand the
dynamics, as studied experimentally in Ref.~\cite{Schmitt2014b}.  It
is important also to note that by including inter-mode coherences,
equation~(\ref{eq:2}) does not make the ``secular approximation'',
which discards those bath-induced terms which are time dependent in
the interaction picture.  The secular approximation is often
introduced as a necessary condition for having a completely positive
master equation~\cite{Duemcke1979}. However several recent papers
suggest the secular approximation can lead to incorrect
predictions~\cite{Jeske2014,Joshi2015,TwoBosonPaper}.  As discussed
later, the current model falls into this class: the
\emph{experimentally observed} oscillations of the photon density can
only occur if the molecular emission produces inter-mode coherence ---
an incoherent state would show no beating.  Beyond the secular
approximation, there may be instabilities, particularly at large
$\GM$.  However, for large $\GM$ the Markov approximation
fails~\cite{TwoBosonPaper}.  For the parameters we consider, our
equations are stable.

Rather than explicitly simulating the density matrix $\rhohat$, we
use the master equation above to write coupled equations of motion for
the (Hermitian) photon correlation matrix $[\vec{n}]_{m,m^\prime} =
\langle \hat{a}^\dagger_m \hat{a}^{}_{m^\prime} \rangle$, and the
coarse-grained excitation density, $f(\vec{r}) = \sum_i
\delta(\vec{r}-\vec{r}_i) \langle \hat{\sigma}^+_i \hat{\sigma}^-_i
\rangle$.  Within the semiclassical approximation~\cite{Kirton2015}
$[\vec{n}]_{m,m^\prime}$ and $f(\vec{r})$ obey a closed set of
equations.  The semiclassical approximation means that we
neglect correlations between the state of the
photons and the dye molecules, so that expectations of products of
operators can be replaced by products of expectations. 
These semiclassical equations can be
written in a particularly compact form by defining a number of other
quantities.  We define the matrices $[\vec{K}^{}_\pm]_{m,m^\prime}
\equiv \delta_{m,m^\prime} K(\pm \delta_m)$, the mode function matrix
$[\vec{\Psi}(\vec{r})]_{m,m^\prime} \equiv \psi_m(\vec{r})
\psi_{m^\prime}(\vec{r})$, the overlap matrix $\vec{f} \equiv \int d^d
\vec{r} f(\vec{r}) \vec{\Psi}(\vec{r})$ and $[\vec{h}]_{m,m^\prime}
\equiv \delta_{m,m^\prime}(i \omega_m - \kappa)$.  We thus write the
equations:
\begin{align}
  \label{eq:3}
  \partial_t \vec{n} = 
  \vec{h} \vec{n}
  + \vec{f} \rho_0 \vec{K}_-(\vec{n}+\mathbbm{1})
  + (\vec{f}-\mathbbm{1}) \rho_0 \vec{K}_+^\dagger \vec{n}
  + \text{H.c.},
  \\
  \label{eq:4}
  \partial_t f(\vec{r}) =
  - \Gdn^{\text{tot}}(\vec n, \vec{r}) f(\vec{r})
  + \Gup^{\text{tot}}(\vec n, \vec{r}) (1-f(\vec{r})),
\end{align}
where $\rho_0$ is the density of molecules.  Note that in the first
equation, while $\vec{n}, \vec{f}$ are Hermitian, the matrices
$\vec{h}, \vec{K}_\pm$ are not.  In the equation for the excitation
density $f(\vec r)$, the total absorption and emission rates are:
\begin{align}
 \Gup^{\text{tot}}(\vec n,\vec{r}) &= \Gup(\vec{r}) + 2 \Re \left( \text{Tr}
  \left[ \vec{\Psi}(\vec{r}) \vec{n} \vec{K}^{}_+ \right] \right) 
\\
\Gdn^{\text{tot}}(\vec n, \vec{r}) &= \Gdn + 2 \Re \left( \text{Tr} \left[
    \vec{\Psi}(\vec{r}) \vec{K}^{}_- (\vec{n} + \vec{1}) \right]
\right),
\end{align}
incorporating also stimulated emission to and absorption from the
cavity modes.  To simulate these equations numerically,
we discretize $f(\vec{r})$ on a grid of $N_x$ spatial points, and
then use an adaptive time-step Runge-Kutta approach to evolve
the coupled equations.

From these equations, we are typically interested in deriving
quantities such as the photon spectrum, and the photon density profile
$I(\vec{r})$ which is our focus in this paper. These quantities can be
directly extracted from the photon correlation matrix.  The spectrum
is given by the diagonal elements $[\vec{n}]_{m,m}$ and the photon
density by $$I(\vec{r}) = \sum_{m,m^\prime} \psi^\ast_m(\vec{r})
\psi^{}_{m^\prime}(\vec{r}) [\vec{n}]_{m,m^\prime}.$$ Simulating the
full state of the system would thus require solving $N_m^2 + N_x$
coupled differential equations.  One can reduce this requirement by
noting that elements of $[\vec{n}]_{m,m^\prime}$ that are far from the
diagonal are however very small.  As discussed earlier, the value of
$N_m$ required is however relatively large; and even with reduction to
terms with small $|m-m^\prime|$, we find it is only feasible to
simulate short-time transient behavior, and only in one spatial
dimension.
The results presented later involved $150$ps of simulated time
requiring four hours of computer time, while the timescale to reach steady
state is of the order of microseconds.

\subsection{Diagonal approximation}
\label{sec:diag-appr}

As noted at the end of the last section, the full equations are too
computationally costly to allow numerical exploration of how the
steady state profile depends on control parameters.  To overcome this
limitation, we introduce here the ``diagonal approximation'', which
is accurate as long as coherence between photon modes is small, and
allows numerical exploration of the steady state.  This
corresponds to using the molecule-photon interaction 
 $
\mathcal{M}_{\text{int}}^{}[\rhohat] =
\mathcal{M}_{\text{int}}^{\text{diag}}[\rhohat]$ 
with
\begin{multline}
  \mathcal{M}_{\text{int}}^{\text{diag}}[\rhohat]
  \equiv
  - i [H_\Lambda, \rhohat]
  +\frac{1}{2}\sum_{m,i} {|\psi_m(\vec{r}_i)|^2}  \times
  \\
  \biggl(
    \Gamma(\delta_m) \mathcal{L}[\hat{a}^{}_m \hat{\sigma}_i^+,\rho]
    +
    \Gamma(-\delta_m) \mathcal{L}[\hat{a}^{\dagger}_m \hat{\sigma}_i^-,\rho]
  \biggr)
\end{multline}
where $\Gamma(\pm\delta)\equiv 2 \Re[K(\pm\delta)]$ are the
absorption(emission) rates, and $H_\Lambda$ is a Lamb shift from the
imaginary part of $K(\pm\delta)$.  This Lamb shift will however be
irrelevant for the equations of motion as discussed next.

In this approximation, we may write closed equations for the
populations of the photon modes, neglecting coherences, and thus have
only $N_m+N_x$ equations.  Denoting the diagonal elements of the
correlation matrix as $n_m = [\vec{n}]_{m,m}$, and the diagonal
overlap elements as $f_m \equiv [\vec{f}]_{m,m} = \int d^d{\vec{r}}
f(\vec{r}) |\psi_m(\vec{r})|^2$, these coupled equations take the
form:
\begin{align}
  \label{eq:5}
  \partial_t n_m &= \moldense \Gamma(-\delta_m)  f_m (n_m+1)
  \nonumber\\&\qquad\qquad
  - \left[ \kappa + \moldense \Gamma(\delta_m)  (1-f_m) \right] n_m, \\
  \label{eq:6}
  \partial_t f(\vec{r}) &=
  - \Gdn^{\text{tot}}(\{n_m\}, \vec{r}) f(\vec{r})
  + \Gup^{\text{tot}}(\{n_m\}, \vec{r}) (1-f(\vec{r})).
\end{align}
Note here that the second equation, Eq.~(\ref{eq:6}), is identical
to that seen previously, however the
total molecular excitation and decay rates are now written in
terms of the diagonal populations are:
\begin{align}
  \label{eq:7}
  \Gdn^{\text{tot}}(\{n_m\}, \vec{r}) &= \Gdn
  +\sum_m |\psi_m(\vec{r})|^2 \Gamma(-\delta_m)(n_m+1),
  \\
  \label{eq:8}
  \Gup^{\text{tot}}(\{n_m\}, \vec{r}) &= \Gup(\vec{r})
  +\sum_m |\psi_m(\vec{r})|^2 \Gamma(\delta_m)n_m.
\end{align}

In all the equations above we have kept the spatial dimension general,
writing generic wavefunctions $\psi_m(\vec{r})$.
The experiments are in two dimensions, in which case the mode
functions should take the form:
\begin{displaymath}
  \psi_m(\vec r) = 
  \frac{%
    {H}_{m_x}\left(\frac{x}{\lho}\right)
    {H}_{m_y}\left(\frac{y}{\lho}\right)
    e^{-r^2/2 \lho^2}}{\lho\sqrt{\pi 2^{m_x+m_y} m_x! m_y!}},
\end{displaymath}
where we take $m=(m_x,m_y)$ as a combined index and $H_m(x)$ is the
$m$th Hermite polynomial.  In certain cases, it is possible to
efficiently find the steady states in two dimensions, and where this
is possible, we follow this approach.  However, when directly solving
the equations of motion, it is intractable to keep all two dimensional
modes with energies $\hbar \omega < k_BT$, and so in some cases below
we instead restrict to one dimension, for which:
\begin{displaymath}
  \psi_m(x) = 
  \frac{1}{\sqrt{\lho\sqrt{\pi} 2^{m} m!}}
  {H}_{m}\left(\frac{x}{\lho}\right)
  e^{-x^2/2 \lho^2}.
\end{displaymath}
In the following we will present analytic results for general
dimension $d$, and specify $d=1$ or $d=2$ for the numerical results.

\section{Steady state}
\label{sec:steady-state}

In the following we  explore the consequences of a finite size
Gaussian pump spot, 
$$\Gup(\vec{r}) = \frac{\Gupint}{(2 \pi \sigma_P^2)^{d/2}}
\exp\left[-\frac{(\vec{r}-\vec{r}_P)^2}{2 \sigma_P^2}\right],$$
where $\Gupint$ is the integrated  intensity, $\sigma_P$ the spot
size, $\vec{r}_P$ the offset, and $d$ the dimension.  Note that 
$\Gupint$ has dimensions of $[T]^{-1} [L]^d$.  Similarly,
since $\Gamma(\pm \delta)$ are multiplied
by $\rho_0$ or $|\psi_m(\vec{r})|^2$, this means $\GM$ also
has dimensions $[T]^{-1} [L]^d$.
We measure all
lengths in units of the  oscillator length
$\lho$ of the harmonic trap potential, and measure all ($d$-dimensional) 
densities in units of $\lho^{-d}$.  For comparison to
Ref.~\cite{Marelic2015}, in the first part of this paper  we set
$\vec{r}_P=0$.  

\subsection{Far below threshold}
\label{sec:far-below-threshold}

\begin{figure}[thpb]
  \centering
  \includegraphics[width=3.2in]{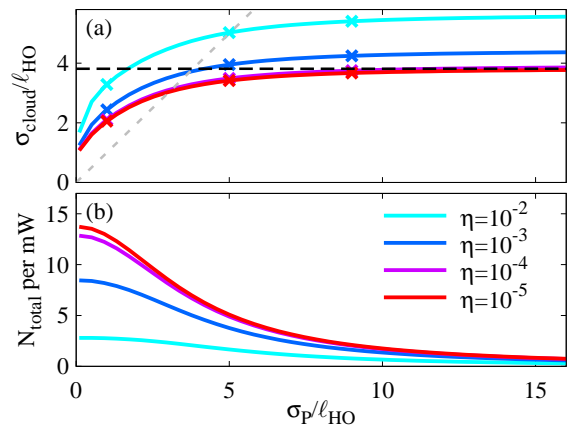}
  \caption{(Color online) (a) Photon cloud size and (b) photon number
    (per unit power) vs pump-spot size, for various $\eta\equiv\kappa/
    \moldense \GM$.  Note that in panel (a), the lines for the two
      smallest values of $\eta$ lie on top of each other.
    Plotted for $d=2$, far below threshold using the
    closed-form solution in Eq.~(\ref{eq:9}) with $\omega_c=3200$THz.
    Dashed lines in (a) show thermal size $\sigma_T$ (see text) and
    pump size $\sigma_P$.  The points marked by the symbols correspond
    to the points for which cross-sections are shown in
    Fig.~\ref{fig:below-thresh-profile}. }
  \label{fig:below-thresh}
\end{figure}

Far below threshold, when $\Gup(\vec{r}=0) \ll
\Gdn$,
both $n_m$ and $f(\vec{r})$ are small, so the steady state 
of Eq.~(\ref{eq:5},\ref{eq:4}) is
\begin{equation}
  \label{eq:9}
  f(\vec{r}) \simeq
  \frac{\Gup(\vec{r})}{\Gdn^{\text{tot}}(\{n_m=0\},\vec{r})},
  \quad
  n_m \simeq f_m \frac{\Gamma(-\delta_m) }{\Gamma(\delta_m) + \kappa/\moldense}.
\end{equation}
Note that in the denominator of the expression for $f(\vec{r})$ we
have not written $\Gdn$, but rather $\Gdn^{\text{tot}}(\{n_m=0\},\vec{r})$,
which includes also the spontaneous emission into empty cavity modes.
However, for relevant parameters (see below), the cavity mode contribution to
$\Gdn^{\text{tot}}(\{n_m=0\},\vec{r})$ is small, so $f(\vec{r}) \simeq
\Gup(\vec{r})/\Gdn$, and the overlaps $f_m$ depend on the pump shape.  In
this limit the shape of $I(\vec{r})$
depends only on the shape of the pump, the normalized spectrum
$\Gamma(\pm\delta)/\GM$, and the dimensionless parameter
$\eta \equiv \kappa/\moldense\GM$.  
If $\eta \ll 1$ and the KS relation is obeyed then:
$n_m =
f_m e^{-\beta \delta_m}$.  If  one also has $\sigma_P \gg \lho$, then
$f_m$ is independent of $m$, 
and so there is a thermal photon distribution leading
to a thermal photon cloud profile:
\begin{displaymath}
  I(\vec{r}) 
  \propto 
  \sum_m e^{-\beta \delta_m} |\psi_m(\vec{r})|^2
  \propto 
  \exp\left( - \frac{r^2}{2 \sigma_T^2}\right),
\end{displaymath}
with $\sigma_T = l_{\text{HO}}/\sqrt{2 \tanh(\beta \epsilon/2)}$,
which recovers the classical thermal cloud size if $k_BT \gg
\epsilon$, see Fig.~\ref{fig:below-thresh}(a).  

\begin{figure}[thpb]
  \centering
  \includegraphics[width=3.2in]{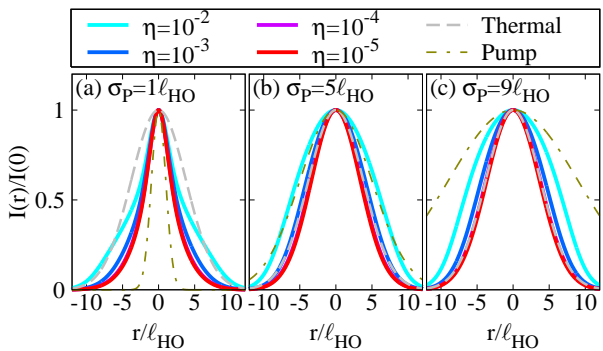}
  \caption{(Color online) Photon cloud profile far below threshold for
    various $\eta\equiv\kappa/ \moldense \GM$.  Note that in all
    panels, the lines for two smallest values of $\eta$ are
    indistinguishable.  The gray dashed line indicates the equilibrium
    profile, and the dash-dotted line indicates the profile of the
    pump spot.  Plotted for $\sigma_P/\lho=1, 5, 9$ respectively (a-c)
    and other parameters as in Fig.~\ref{fig:below-thresh}.}
  \label{fig:below-thresh-profile}
\end{figure}

Thermalization fails for small $\sigma_P$ or large $\eta$.  This can
be seen by looking at the actual cloud profiles, as shown in
Fig.~\ref{fig:below-thresh-profile}.  At small $\sigma_P$ this failure
is due to the mode dependence of $f_m$: A small pump spot populates
only the low order photon modes, leading to an unnaturally small
(i.e.\ cold) photon cloud, even when $\eta \ll 1$ (see
Fig.~\ref{fig:below-thresh-profile}(a)).  Note that for this
non-thermal distribution to occur in this limit, the presence of the
non-cavity decay rate $\Gdn$ is crucial: if both sources of loss
$\Gdn$ and $\kappa$ are small, repeated absorption and re-emission of
photons will occur, producing a thermal distribution.  For large
$\eta$ thermalization fails because $n_m \simeq f_m
\Gamma(-\delta_m)/\kappa \rho_0$, and so no Boltzmann factor arises.
In Fig.~\ref{fig:below-thresh}(a) this gives a photon cloud which is
larger than $\sigma_T$ (see outermost (cyan) line in all panels
of Fig.~\ref{fig:below-thresh-profile}).

Figure~\ref{fig:below-thresh}(b) shows the total photon number (per
unit of incident power) vs pump spot size.  For large spots, the 
 number falls off as $(\sigma_P)^{-d}$.  This is because for all
modes $m$ with extent much smaller than $\sigma_P$, one may approximate
\begin{equation}
  \label{eq:10}
f_m \simeq \frac{\Gup(\vec{r}=0)}{\Gdn} =\frac{\Gup^{\text{int}}}{\Gdn(\sqrt{2\pi}
  \sigma_P)^d},
\end{equation}
thus giving $N_{\text{total}}/\Gup^{\text{int}} \equiv \sum_m f_m  e^{-\beta \delta_m}
/\Gup^{\text{int}} \propto \sigma_P^{-d}$.
In contrast, for small spots, the number saturates; here $\sigma_P$ is
much smaller than the extent of the relevant modes and so $f_m \simeq
|\psi_m(\vec{r}=0)|^2 \Gup^{\text{int}}/\Gdn$, independent of
$\sigma_P$.  This expression clearly means that the occupation of all
odd modes (which have a node at $\vec{r}=0$) will vanish.  Even modes
however have a value for $\psi_m(\vec{r}=0)$, and so $N_{\text{total}}$
clearly saturates at a finite value.  Once $\sigma_P \ll \lho$, this
overlap with even modes becomes independent of $\sigma_P$, leading to
the saturation observed in Fig.~\ref{fig:below-thresh}.  In
experiment~\cite{Marelic2015}, the total photon number initially
increases with spot size, an effect not seen here.  Such a discrepancy
could perhaps arise if the coupling of small pump spots into the
cavity is less efficient due to some aspect of the pumping
optics:  given the above arguments about overlaps for
small pumping spots, it is hard to explain such a low efficiency
when considering purely light trapped inside the cavity.

The behavior in Fig.~\ref{fig:below-thresh}(a) for $\eta \simeq
10^{-3}$ is very similar to the experimental results of
Ref.~\cite{Marelic2015}.  Using other known parameters of this
experiment, $\rho_0 \simeq 10^8 \lho^{-2}$, and $\kappa=500$MHz, this
gives $\GM=5$kHz$\lho^2$. 
We use these parameter values below unless otherwise stated.
In order to verify the assumption made earlier that
we may replace 
$\Gdn^{\text{tot}}(\{n_m=0\},\vec{r}) \simeq \Gdn$, we use these values
to estimate the effect of loss into empty cavity modes.
Comparing the rate of emission into
the lowest cavity mode,
$\Gamma(-\delta_m)|\psi_0(\vec{r}=0)|^2 < \GM / (\sqrt{\pi} \lho^2)
\simeq 2.8$kHz,
to the observed background decay rate $\Gdn \simeq 250$MHz, this implies
that even if the first $1000$ cavity modes contributed to the emission
equally, the cavity-mediated contribution would be far smaller than 
the background.  The contribution of high order cavity modes however falls off
due both to the overlap $\psi_m(\vec r)$, and the eventual
decay of $\Gamma(-\delta_m)$ at large $m$.  Thus, the assumptions made at
the start of this section are indeed justified.

\begin{figure}[htpb]
  \centering
  \includegraphics[width=3.2in]{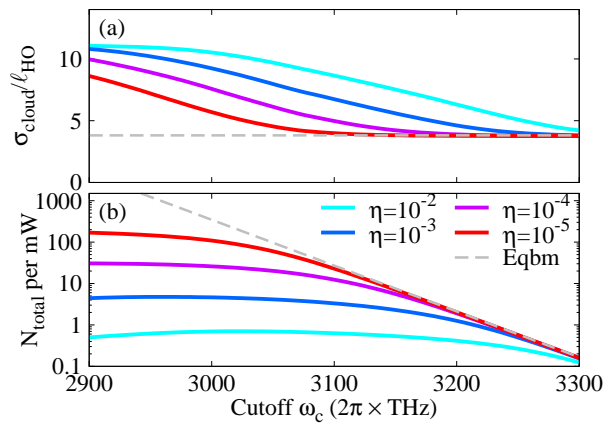}
  \caption{(Color online) (a) Photon cloud size and (b) photon number
    (per unit power) vs cavity cutoff $\omega_c$ for various
    $\eta\equiv\kappa/ \moldense \GM$.  For comparison to later
    figures, this is plotted for d=1.  Spot size $\sigma_P=16\lho$;
    all other parameters as in Fig.~\ref{fig:below-thresh}.}
  \label{fig:low-power-vs-cutoff}
\end{figure}

So far in this section we have explored dependence on the properties
of the pump spot. Another relatively easy parameter to tune is the
cavity cutoff frequency $\omega_c$.  Indeed, as discussed extensively
by~\cite{Schmitt2014b}, tuning this parameter can be used to control
the degree of thermalization.  A large value of $\omega_c$ will
enhance reabsorption of cavity and thus lead to thermalization, while
a smaller value reduces reabsorption and prevents equilibration.
Later in this paper we discuss this behavior at and above threshold.
Figure~\ref{fig:low-power-vs-cutoff} shows the effect of cutoff
frequency on properties far below threshold. One can clearly see in
Fig.~\ref{fig:low-power-vs-cutoff}(a) that at large $\omega_c$, the
photon cloud size approaches the equilibrium cloud size.  Similarly,
considering the total number of photons, one sees that at large
$\omega_c$ the behavior asymptotically approaches the equilibrium
result $N = \sum_m f_m e^{-\beta \delta_m}$ indicated by the gray
dashed line (where $f_m$ is given by Eq.~(\ref{eq:10}) as we consider
a large cloud size, $\sigma_P \gg \lho$).  At smaller $\omega_c$, the
photon number decreases as $N \simeq \sum_m f_m\Gamma(-\delta_m)
\moldense/\kappa$, hence the strong dependence seen upon the value of
$\eta = \kappa/\moldense \GM$.

\subsection{Threshold pump power}
\label{sec:behav-at-thresh}

\begin{figure}[htpb]
  \centering
  \includegraphics[width=3.2in]{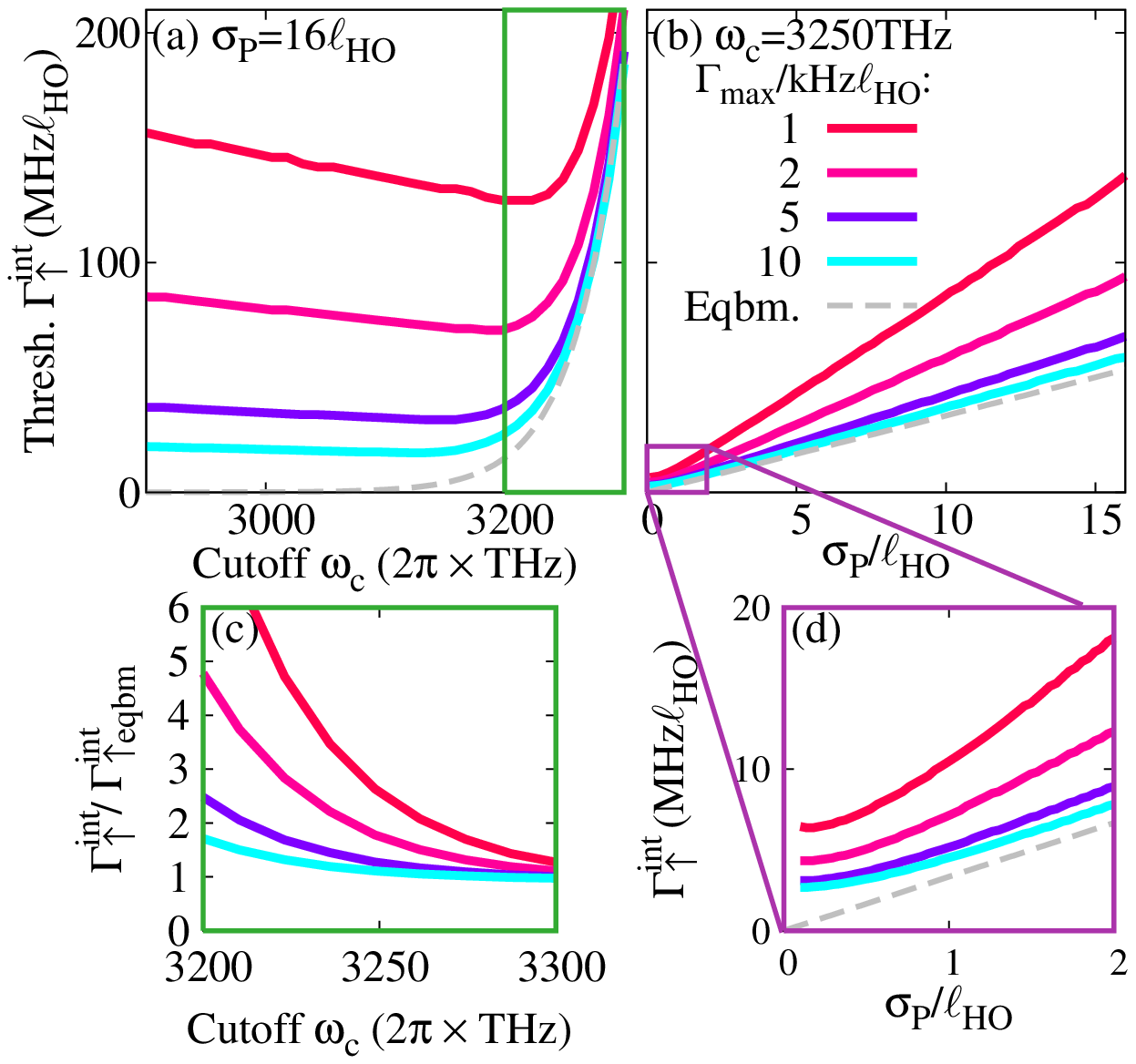}
  \caption{(Color online) Threshold (integrated) pump power,
    calculated in $d=1$ for various values of $\GM$.  (a) vs cavity
    cutoff, $\omega_c$, (b) vs pump spot size.  Simulations performed
    in one spatial dimension with $N_m=200$ photon modes, and
    $N_x=300$ spatial grid points.  The dashed line shows the
    equilibrium result $\Gupint \equiv \sqrt{2\pi} \sigma_P \Gup(r=0)=
    \sqrt{2\pi} \sigma_P \Gdn e^{\beta \delta_c}$.  At large spot
    sizes, the threshold power increases as $(\sigma_P)^d$ and so is
    linear for this 1D simulation.  Panel (c) shows the ratio of
    threshold power vs equilibrium threshold power for the region
    highlighted in panel (a), illustrating the asymptotic approach to
    equilibrium.  Panel (d) shows an enlarged region from panel (b) as
    indicated.}
  \label{fig:threshold}
\end{figure}

We next turn to the behavior at threshold, and explore how this
depends on the pump size.  As with any finite size system, the
threshold is not perfectly sharp; for definiteness we use the same
threshold condition defined in Refs.~\cite{Kirton2013b,Kirton2015}.
Figure~\ref{fig:threshold} shows the threshold value of $\Gupint$ vs
cavity cutoff, $\omega_c$, and vs pump spot size, $\sigma_P$.  These
calculations, time-evolving Eqs.~(\ref{eq:5}--\ref{eq:8}) numerically,
are computationally expensive in $d=2$ so we consider $d=1$ from
hereon.  In equilibrium the threshold condition is $\Gup(\vec{r}=0) =
\Gdn e^{\beta \delta_c}$, where $\delta_c = \omega_c-\wzpl$
(see~\cite{Kirton2013b}).  This condition has a simple meaning: it
identifies when the effective chemical potential of the molecules,
$\mu_{\text{eff}}(\vec{r}) = \wzpl + k_BT
\ln[\Gup(\vec{r})/\Gdn]$ reaches the lowest photon
mode~\cite{Kirton2015}, $\omega_c$.

As has been discussed
previously~\cite{Marthaler2011,Kirton2013b,Kirton2015}, it is notable
that the lasing transition, normally associated with inversion, can be
described here as corresponding to a thermal distribution with a
positive temperature and chemical potential $\mu_{\text{eff}} <
\wzpl$.  The fundamental reason why electronic inversion is not
required here is the different rates of absorption and emission,
$\Gamma(\pm \delta)$.   Net inversion of the electronic state is
\emph{normally} required for lasing because absorption and emission
coefficients match, so net gain requires an inverted population.
Here, for modes with $\delta = \omega-\wzpl < 0$, we have
$\Gamma(-\delta) > \Gamma(\delta)$, and so for these modes, emission
exceeds absorption even without electronic inversion.  If one
considers the microscopic ro-vibronic levels of the molecule, there is
inversion between the lowest ro-vibrational level of the electronic
excited state,  and the higher ro-vibrational levels of the electronic
excited states, and it is these transitions that have 
net gain~\cite{Schafer1990}. However, in our model, the fast dynamics
of the ro-vibrational levels have been adiabatically eliminated.

The thermal equilibrium prediction of threshold at
$\mu_\text{eff}(\vec{r}=0)=\omega_c$ is shown as the dashed line in
Fig.~\ref{fig:threshold}.  The actual threshold in
Fig.~\ref{fig:threshold}(a) is however non-monotonic.  At large
$\omega_c$ the system is thermal, and so threshold increases
exponentially with $\omega_c$.  Figure \ref{fig:threshold}(c)
illustrates the asymptotic approach to the equilibrium behaviour by
plotting the ratio $\Gupint/(\sqrt{2\pi} \sigma_P \Gdn e^{\beta
  \delta_c})$, which approaches $1$ at large $\omega_c$.  At small
$\omega_c$ the absorption and emission rates are too small to compete
with cavity loss, and so the threshold pump increases.  Such
non-monotonic dependence has been seen
experimentally~\cite{Marelic2015}.  The minimum of threshold becomes
more pronounced as one increases the cavity loss rate $\kappa$ or
decreases the peak emission rate $\GM$.

In Figure~\ref{fig:threshold}(b), we see that in $d=1$,
$\Gup^{\text{int}} \propto \sigma_P$ at threshold, except for small
spot sizes where it saturates (see the enlarged region in
Fig.~\ref{fig:threshold}(d)).  From the asymptotic form of the
equations it is straightforward to see that in $d$ dimensions this
result becomes $\Gup^{\text{int}} \propto (\sigma_P)^d$.  Such a
dependence on spot size occurs because threshold is reached first at
the trap center, where $\mu_{\text{eff}}(\vec{r})$ is greatest.  As
such, it is the peak pump power $\propto \Gup^{\text{int}}
\sigma_P^{-d}$ that enters the threshold condition.  It is however
important to note that the simple power law arises only for large
enough spot sizes, whereas for spot sizes comparable to the harmonic
oscillator length, saturation of the critical integrated power occurs.
In the (two dimensional) experiment of Ref.~\cite{Marelic2015} a
phenomenological power law with exponent $\sim 1.5$ was extracted from
a least squares plot to data on a log-log scale over one decade of
spot size.

\begin{figure}[htpb]
  \centering
  \includegraphics[width=3.2in]{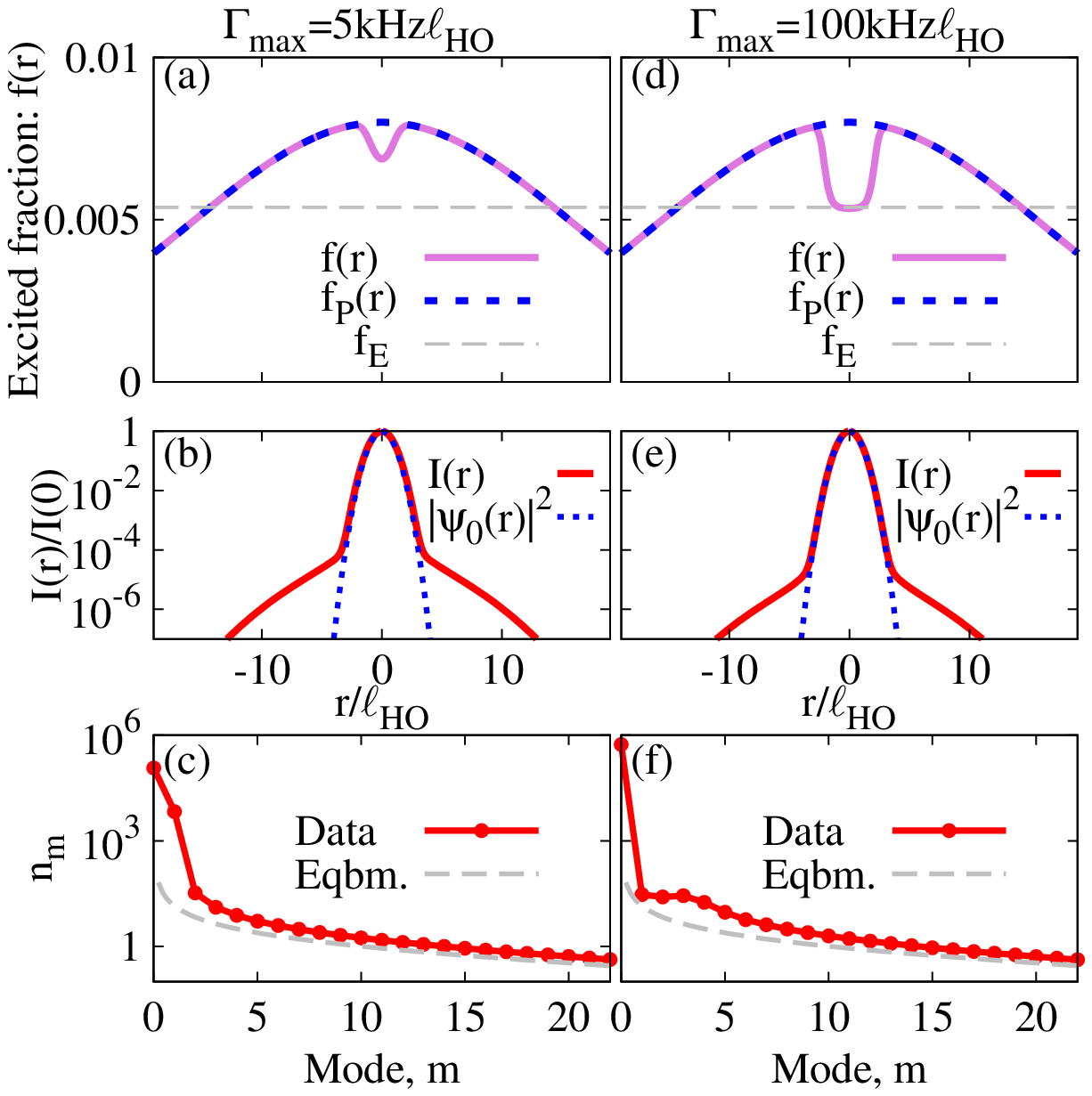}
  \caption{(Color online) Clamping of the gain profile for
    sufficiently large $\GM$ (in $d=1$).  Panels (a,b,c) show
    $\GM=5$kHz$\lho$, panels (d,e,f) $\GM=100$kHz$\lho$.  Top panels
    show the gain profile $f(\vec{r})$ (solid), the value set by the
    pump, $f_{P}(\vec{r}) = \Gup(\vec{r})/(\Gdn+\Gup(\vec{r}))$ (short
    dashed), and the clamped value $f_{E}=[e^{-\beta\delta_c}+1]^{-1}$
    for the cavity cutoff $\omega_c=3200$THz.  Middle panels show the
    (normalized) photon density plotted on a logarithmic scale.  For
    comparison the blue dashed line shows the profile of the ground
    mode $|\psi_0(r)|^2$.  Bottom panels show the mode populations
    $n_m$ in comparison to an equilibrium Bose-Einstein distribution,
    demonstrating multimode condensation.  Simulations performed in
    one spatial dimension with $N_m=200$ photon modes, and $N_x=300$
    spatial grid points.  }
  \label{fig:saturation}
\end{figure}

\subsection{Above threshold --- transverse hole burning}
\label{sec:above-thresh-transv}

Once threshold is reached, in equilibrium, the chemical potential
locks at $\mu_{\text{eff}}=\omega_c$. This means that the ``gain
profile'' $f(\vec{r})$, i.e.\ the fraction of excited molecules, must
saturate, $f(\vec{r}) \le f_E = [e^{-\beta \delta_c} +1]^{-1}$.  Such
a saturation of $f(\vec r)$ is also expected for a laser, and is known
in that context as gain clamping.  As noted above, because of the
different rates $\Gamma(\pm\delta)$, no electronic inversion is
required for lasing, and so net gain exists even though $f(\vec{r}) <
1$. As such, for a photon BEC, the laser concept of gain clamping and
the thermal concept of chemical potential locking are the same.  Gain
clamping or chemical equilibrium also imply that $f(\vec{r})$ should
become uniform at threshold, and we next turn to explore if and how
this occurs.  Figure~\ref{fig:saturation} shows $f(\vec{r})$ slightly
above threshold for two values of $\GM$.  At $\GM=100$kHz$\lho$,
clamping is seen near the trap center, but for $\GM=5$kHz$\lho$ it is
absent.  The dependence on $\GM$ follows from the steady state
result, $$f(\vec{r}) =
\frac{{\Gup^{\text{tot}}(\{n_m\},\vec{r})}}{\Gdn^{\text{tot}}(\{n_m\},\vec{r})+{\Gup^{\text{tot}}(\{n_m\},\vec{r})}}$$
and the form of
$\Gamma_{\uparrow,\downarrow}^{\text{tot}}(\{n_m\},\vec{r})$ in
Eq.~(\ref{eq:7},\ref{eq:8}). If $n_m$ is a Bose-Einstein distribution
with chemical potential $\mu$ and $\Gamma(\delta)$ obeys the KS
relation then $$\Gamma(-\delta_m) (n_m+1) = \Gamma(\delta_m) n_m
e^{-\beta \mu}.$$ This means that if both the following are obeyed:
\begin{align*}
&\sum_m |\psi_m(\vec{r})|^2 \Gamma(-\delta_m)(n_m+1) \gg \Gdn, \\
&\sum_m |\psi_m(\vec{r})|^2 \Gamma(\delta_m)n_m \gg \Gup(\vec{r}),
\end{align*}
then one has an uniform gain profile $f(\vec{r}) = f_E$.  We thus see
that gain clamping requires large $\Gamma(\pm\delta)$.  Moreover,
since the condensed mode(s) are concentrated at the trap center, gain
clamping is spatially restricted, as seen in
Fig.~\ref{fig:saturation}(d).  In the terminology of lasers, this is
analogous spatial hole burning~\cite{Siegman1986}.  In standard
laser resonators, hole burning is discussed in terms of competition among
different longitudinal modes, as typically laser resonators are much
longer than the wavelength of light, but designed to support few
transverse modes.  In the photon BEC the situation is opposite: the
microcavity supports only one longitudinal mode nearly resonant with
the gain medium, but many transverse modes.  As such, one has
``transverse spatial hole burning'', leading to patterns in the gain
medium as a function of the transverse coordinate, as opposed to the
more standard patterns along the cavity axis.

As the gain clamping is imperfect, other modes may reach threshold
leading to multimode condensation.  In Fig.~\ref{fig:saturation} 
panels (c,f) show the mode populations $n_m$, demonstrating that several
modes are macroscopically occupied.  For values of $\GM$ larger than
those shown, multimode condensation is suppressed.  Both the
inhomogeneous $f(\vec{r})$ and multimode condensation are signatures
of imperfect thermal equilibrium.  An interesting question for future
work is how such hole burning might be used to engineer the photon
condensate profile.

\section{Dynamics}
\label{sec:dynamics}

Having discussed the effects of the pump-profile on steady-state
properties, we now turn to consider dynamics, and the transient
response following a pump pulse.  The calculations in this section are
motivated particularly by the work of \citet{Schmitt2014b}, who
studied the dynamics of the photon condensate after an off-center pump
pulse and observed oscillations of the photon condensate.  These
oscillations correspond to transverse motion of a photon wavepacket in
the effective harmonic trap potential of the mirrors.
\citet{Schmitt2014b} showed in particular that for a higher frequency
cavity cutoff $\omega_c$ (closer to the peak of the molecular emission
and absorption spectrum), thermalization occurred, while for a lower
cutoff, oscillations persisted to later times.  \citet{Schmitt2014b}
also provided a theoretical discussion of their results, making use of
a set of semiclassical equations for quantities $n_m(\vec r)$,
i.e. spatially dependent population of a given mode.  Such a quantity
does not appear within the model discussed above: a given mode has a
given spatial profile $|\psi_m(\vec r)|^2$, and as discussed earlier,
this means that within the diagonal model one cannot have asymmetric
distributions, since the mode profiles $|\psi_m(\vec r)|^2$ are all
even. Our aim in this section is therefore to show that our model can
reproduce the behavior seen by \citet{Schmitt2014b}, while considering
the full covariance matrix $[\vec{n}]_{m,m^\prime}$, or equivalently
its spatial representation $\vec{n}(\vec r, \vec r^\prime) =
\sum_{m,m^\prime} [\vec{n}]_{m,m^\prime} \psi_m(\vec r) \psi_m(\vec
r^\prime)$.

To explore both thermalization and oscillations, 
it is crucial to use the model as presented in
Eq.~(\ref{eq:2}), i.e.\ without the secular approximation.
This can be seen from quite general arguments.
Firstly, in order to describe an off-center photon pulse, we must
allow emission into wavepackets, not just populations of eigenstates,
since $|\psi_m(\vec r)|^2$ is symmetric for all modes, and so cross
terms $\psi_m^\ast(\vec r) \psi_{m^\prime}(\vec r)$ are crucial to give
an off center photon distribution $I(\vec r)$.  Including cross terms
is also crucial in order to describe oscillating wavepackets since
time dependence occurs via beating between modes.  This is
incompatible with the standard approach of secularizing master
equations to produce a Lindblad form: secularization is appropriate
if one may  assume
that any cross terms between different modes oscillate fast, and so
should be removed.  The result is an equation which can only produce
populations of modes.  Physically it is however clear that the beating
between modes is not a fast process to be eliminated, but a process on
timescales comparable to emission and absorption.  

Naively, including cross terms between photon modes
might suggest an alternate phenomenological equation which is of Lindblad form:
\begin{multline}
  \mathcal{M}^{}_{\text{int}} [\rhohat] = 
  \Gamma(-\delta)
  \mathcal{L}\left[ \sum_{m,i} \psi^\ast_m(\vec r_i) \hat{a}^\dagger_m
  \hat{\sigma}_i^-, \rhohat \right] 
  \\+ \Gamma(\delta) \mathcal{L}\left[ \sum_{m,i}
  \psi^{}_m(\vec r_i) \hat{a}^{}_m \hat{\sigma}_i^+, \rhohat \right].
\label{eq:11}
\end{multline}
However, such a form is not able to describe thermalization, and its
dependence on cutoff wavelength.  As discussed earlier, such
thermalization relies on the fact that emission and absorption rates
on the detuning of a given mode, $m$, but by its form, Eq.~(\ref{eq:11}) 
has rates independent of mode.  Modeling both the emission into
wavepackets (i.e.\ inclusion of cross terms), and the
mode-frequency-dependence of emission rates requires an equation of
the form of Eq.~(\ref{eq:2}).

\begin{figure}[!htpb]
  \centering
  \includegraphics[width=3.2in]{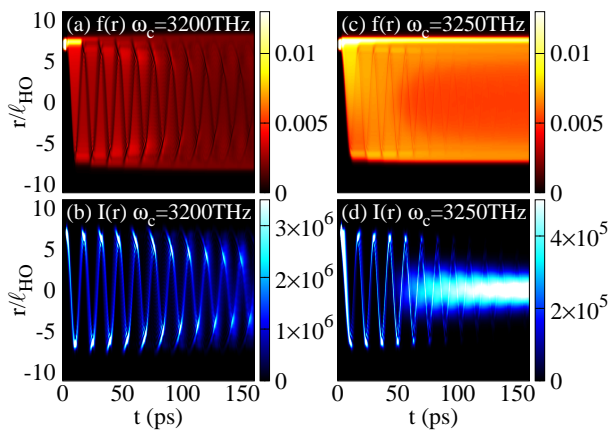}
  \caption{(Color online) Oscillations following an off-center pump pulse (in $d=1$).
    Panels(a,b) are for $\omega_c=3200$THz, where thermalization is
    not sufficient to suppress the oscillations. Panels (c,d) are for
    $\omega_c=3250$THz, showing a transition to a central time-independent 
photon cloud at late times.  Panels
    (a,c) show $f(\vec{r})$, and (b,d) show $I(\vec{r})$.  The pump
    pulse is a Gaussian at $r_P=7\lho$, with width $\sigma_P=0.3\lho$,
    duration duration $5$ps and intensity $\Gupint=24$GHz$\lho$.
    Other parameters; $\kappa=100$MHz, $\Gdn=250$MHz,
    $\GM=3$MHz$\lho$, mode spacing $\epsilon=0.4$THz.  These last two
    parameters are enhanced compared to experiments to reduce
    simulation time.  Simulations performed in one spatial dimension
    with $N_m=180$ photon modes, keeping inter-mode coherences for
    modes with $|m-m^\prime|<40$, and $N_x=300$ spatial grid points.
    This gives $~15,000$ coupled equations, and this requires four
    hours to simulate $150$ps.}
  \label{fig:oscillation}
\end{figure}

Using Eq.~(\ref{eq:3},\ref{eq:6}), we simulate the dynamics following
a short, high intensity, off-center pump pulse.  The resulting photon
density $I(\vec r)$ and excited molecule fraction $f(\vec r)$ are
shown in Fig.~\ref{fig:oscillation}.  The behavior differs according
to the cavity cutoff $\omega_c$ as seen
experimentally~\cite{Schmitt2014b}.  Oscillations occur initially in
both cases, but at late times, they are replaced by a cloud near the
trap center when $\omega_c$ is large enough.  Note that, as seen in
experiment, the switch to the thermal condensate does not occur
through a continuous damping of the amplitude of the oscillations, but
rather through a growing intensity of the central cloud, and decaying
intensity of the oscillating cloud.

\begin{figure}[htpb]
  \centering
  \includegraphics[width=3.2in]{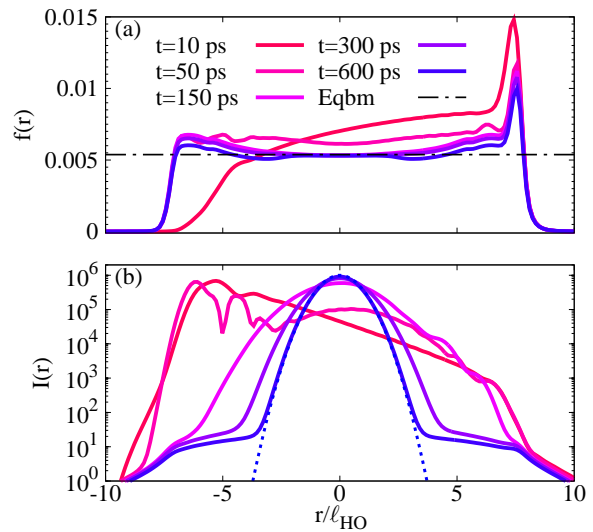}
  \caption{(Color online) Cross sections of
    Fig.~\ref{fig:oscillation}, for $\omega_c=3250$THz.  The gray
    dashed in panel (a) is the equilibrium population of molecules.
    In panel (b) the blue dashed line shows the profile of the ground
    mode $|\psi_0(r)|^2$. }
  \label{fig:thermalization-xs}
\end{figure}

As first shown by \citet{Schmitt2014b}, the
origin of thermalization is that thermalization occurs when
re-absorption of photons leads to a flat gain profile $f(\vec{r})
\simeq f_E$ in the center of the trap.  Our model also reproduces
this behavior, as can be seen from
Fig.~\ref{fig:oscillation}(c), and is also more clearly shown in 
 Fig.~\ref{fig:thermalization-xs} which plots cross sections
of $f(\vec r)$ at various time slices.  

\begin{figure}[htpb]
  \centering
  \includegraphics[width=3.2in]{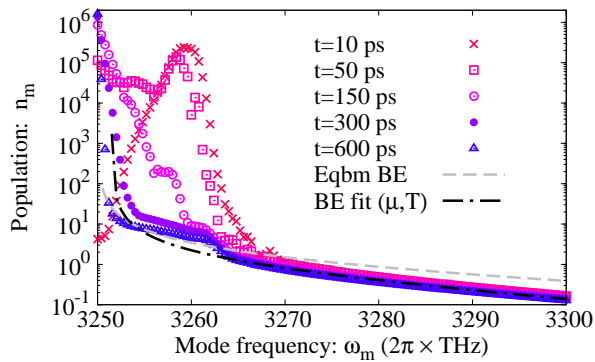}
  \caption{(Color online) Spectrum (i.e.\ diagonal elements of
    correlation matrix, $[\vec{n}]_{m,m}$), plotted at the same times
    as the cross sections in Fig.~\ref{fig:thermalization-xs}, plotted
    for $\omega_c=3250$THz.  The black dash-dotted line in involves
    fitting a Bose-Einstein distribution, the fitting temperature is
    $175$K, and chemical potential $\mu=3251$THz --- see further
    discussion in the text.  The gray dashed line is for comparison
    the Bose-Einstein distribution with the Equilibrium parameters
    $T=300$K, $\mu=3250$THz }
  \label{fig:thermalization}
\end{figure}

Figure~\ref{fig:thermalization} shows how the photon spectrum evolves
with time for $\omega_c=3250$THz, where oscillations disappear at late
times.  As also seen by \cite{Schmitt2014b}, the high energy modes
rapidly reach a steady population, while the lower energy modes evolve
more slowly, and are still evolving even after $600$ps of simulation
time.  Note however that the occupation of the high energy photon
modes does not match the dye temperature.  This is because of the
limited spatial extent of the gain profile $f(\vec r)$ -- this extends
over a range $r/\ell \lesssim 7$, so modes up to $m \lesssim 50$ will
be effectively populated.  This corresponds to $\omega_m = \omega_c +
m \epsilon \lesssim 3270$, higher modes start to be suppressed by
spatial overlap, leading to colder photon distribution, as discussed
earlier.

\section{Discussion and conclusions}
\label{sec:conclusion}

In conclusion, we have presented a theoretical model capable of
describing the spatial dynamics, relaxation, and thermalization of an
inhomogeneously pumped photon condensate.  Using this, we have
reproduced recent experimental results studying the effects of small
pump spots.  Even without photon loss, thermalization can be inhibited
by small spot size.  Our model gives direct access to the gain profile
$f(\vec{r})$, which is hard to access experimentally.  By doing so, we
see the observed behavior at and above threshold is related to gain
clamping and spatial hole burning.

In this paper we have presented results only for Gaussian pump spots
and harmonic trapping potentials.  However, the equations we present
can easily be generalized to other pump profiles or trapping
potentials, e.g. ring-shaped pumps.  As such,
Eqs.~(\ref{eq:3},\ref{eq:4}) or the diagonal approximation,
Eqs.~(\ref{eq:5},\ref{eq:6}), provide a useful model to predict how
the spatial and spectral structure of photon condensates is determined
by the properties of the pump.  As well as the computation framework,
some general principles can be identified from our results.  Far below
threshold, the condensate profile is simply given in terms of the
overlaps $f_m$ between the pump profile and a given trap mode,
allowing one to understand how the cloud will shrink if $f_m$ is
significant only for low-order modes. A ring pump would in contrast
lead to overlaps for a specific range of $m_x, m_y$, and a
corresponding ring-shaped thermal cloud.  Far above threshold, the
picture of gain saturation and spatial transverse hole burning can
provide an intuitive picture of which condensate modes are favored or
suppressed by a particular pump profile, and which profile shapes
favor single or multimode condensation.

Mode competition and spatial pattern formation in driven dissipative
systems have recently prompted significant interest in other contexts,
e.g.\ for random lasers~\cite{Tureci2008,Ge2010}, where the role of
mode competition and the statistics of multimode lasing have been
studied.  Mode competition is the basis of transverse pattern
formation in nonlinear optics~\cite{denz03,staliunas}, and is a prime
example of pattern formation out of equilibrium~\cite{cross93}.  The
model and results presented in this paper provide the foundation to study these
effects in the photon condensate.

As noted in the introduction, for both lasers and condensates, order
parameter equations are widely used to describe the spatial dynamics.
In contrast, our work here is based on solving equations for the
photon density matrix directly.  By solving all elements of the photon
density matrix, we allow for populations of modes in addition to the
condensate mode.  Such a treatment is crucial in reproducing thermal
expectations, particularly below threshold.  In the absence of noise
terms, this is not possible with standard order parameter equations.
However, thermal fluctuations can be incorporated into classical field
methods by adding stochastic noise terms.  This is discussed
extensively in the review article by \citet{Blakie:Dynamics}.  It is
an interesting question for future work to develop a stochastic order
parameter equation that can reproduce the dynamics studied here.

As compared to work on pattern formation in polariton
condensates~\cite{wertz10,Manni2011,Tosi2012,Nelsen2013,Gao2015}, an
advantage of the photon condensate system is that we possess a clear
model of the processes leading to the thermalization, and one which is
readily tractable for spatially extended systems.  For polaritons,
various phenomenological models~\cite{Wouters2010b,Wouters2012a} have
been developed, and attempts to derive models
microscopically~\cite{Haug2014} have been made.  However questions
remain open about the relative role of polariton--polariton scattering
vs.\ scattering with phonons in the
semiconductor~\cite{cao04,doan05:prb}.  In contrast, the models
provided in this paper for the weak-coupling photon condensate are far
simpler, as the effect of the (localized) ro-vibrational modes are well
characterized through the function $\Gamma(\delta)$.  An interesting
question arising from this is to explore how to extend the
treatment presented here to the case of strong coupling with organic
molecules~\cite{Forrest2010,Plumhof2013,Daskalakis2014}.

\acknowledgments{We are very happy to acknowledge stimulating
  discussions with R.~A.~Nyman, J.~Klaers, M.~Weitz and V.~Oganesyan.
  We also wish to thank R.~Nyman for providing the measured absorption
  and luminescence data shown in Fig.~\ref{fig:spec}. The
  authors acknowledge financial support from EPSRC program ``TOPNES''
  (EP/I031014/1) and EPSRC (EP/G004714/2).  JK acknowledges support
  from the Leverhulme Trust (IAF-2014-025). PGK acknowledges support
  from EPSRC (EP/M010910/1).}

\appendix

\section{Extracting $\Gamma(\delta)$ from experimental spectra}
\label{sec:extr-gamm-from}

As discussed in section~\ref{sec:model}, we aim to use the full
spectrum, $\Gamma(\delta)$, extracted from experimental measurements.
This spectrum includes the effects of all ro-vibrational modes automatically.
The experimental measurements provide two functions
$\Gamma_{\text{abs.,exp}}(\omega), \Gamma_{\text{fluor., exp}}(\omega)$,
corresponding to  absorption and fluoresence measurements.
This appendix describes the procedure we use to find a spectrum
consistent with both these measurements and the Kennard--Stepanov
relation.

We first produce a single experimental function $\Gamma(\delta)$, by
averaging the absorption and fluorescence measurements.  This is done
by identifying $\wzpl$ from the mid-point between the peak
absorption and emission, and then shifting and overlapping the
experimental spectrum about these points to produce an averaged
experimental function.  We use a cubic-spline fit to the experimental
data, and construct the function $\Gamma_{\text{exp}}(\delta) =
[\Gamma_{\text{abs.,exp}}(\wzpl+\delta) +
\Gamma_{\text{fluor., exp}}(\wzpl-\delta)]/2$.  This yields a
single function, but not one consistent with the Kennard--Stepanov
relation.  Furthermore, at large negative $\delta$, where
$\Gamma_{\text{exp}}(\delta)$ is small, it falls below a noise floor,
and so the experimental measurements cannot probe the exponentially
small values that must be present to satisfy Kennard--Stepanov.

To address both the above points, we then construct the function:
\begin{equation}
  \Gamma(\delta) =
  \frac{1 + x(\delta)}{2}
  \Gamma_{\text{exp}}(\delta)
  + 
  \frac{1 - x(\delta)}{2}
  \Gamma_{\text{exp}}(-\delta) e^{\beta \delta},
\end{equation}
where $x(\delta)$ interpolates smoothly from $-1$ at large negative $\delta$ to
$+1$ at large positive $\delta$, such that $x(-\delta) = - x(\delta)$.
One may readily check that this ensures $ \Gamma(-\delta) e^{\beta
  \delta} = \Gamma(\delta)$ as required.  The interpolation has the
effect that we use $\Gamma(\delta) \simeq \Gamma_{\text{exp}}(\delta)$
where $\Gamma_{\text{exp}}(\delta)$ is large, and $\Gamma(\delta)
\simeq \Gamma_{\text{exp}}(-\delta) e^{\beta \delta}$ where
$\Gamma_{\text{exp}}(\delta)$ is small and below the noise floor, but
$\Gamma_{\text{exp}}(-\delta)$ is large --- see Fig.~\ref{fig:spec}.

%

\end{document}